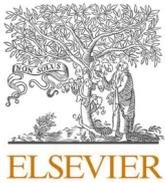
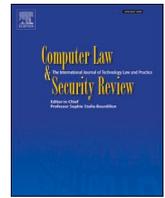

# The algorithmic muse and the public domain: Why copyright's legal philosophy precludes protection for generative AI outputs

Ezieddin Elmahjub [1]

*Associate Professor of Law, Qatar University, Doha, Qatar*

ABSTRACT

Generative AI (GenAI) outputs are not copyrightable. This article argues why. We bypass conventional doctrinal analysis that focuses on black letter law notions of originality and authorship to re-evaluate copyright's foundational philosophy. GenAI fundamentally severs the direct human creative link to expressive form. Traditional theories utilitarian incentive, labor desert and personality fail to provide coherent justification for protection. The public domain constitutes the default baseline for intellectual creations. Those seeking copyright coverage for GenAI outputs bear the burden of proof. Granting copyright to raw GenAI outputs would not only be philosophically unsound but would also trigger an unprecedented enclosure of the digital commons, creating a legal quagmire and stifling future innovation. The paper advocates for a clear distinction: human creative contributions to AI-generated works may warrant protection, but the raw algorithmic output should remain in the public domain.

## 1. Introduction

An art historian prompts an image generator: "A lost sketch by Käthe Kollwitz depicting a mother sheltering her child from a data storm." Potentially very coherent artwork is made in seconds. It is not a collage or a copy, but a novel image, rendered in charcoal strokes that capture Kollwitz's stark anguish. Yet, the work is unsettlingly alien. The "data storm" is represented as a cascade of glyphs no human hand drew, and the mother's expression contains a digital stillness foreign to the artist's known work. The prompt was a simple human request; the output is a complex, machine-mediated creation whose expressive essence is difficult to attribute. This scenario exemplifies the new reality of creative production. GenAI now produces literary, artistic, and audio-visual works, transforming the creative landscape far beyond the assistance of traditional tools. Unlike conventional instruments such as brushes, cameras, or word processors, GenAI actively participates in the generation of outputs, transforming the creative process into a probabilistic collaboration between human intent and algorithmic in-ference. This profound shift compels us to confront an urgent philosophical and legal question: Should copyright subsist in works generated by GenAI? I argue that it should not. My reasoning rests on the foundational philosophy of copyright law: the public domain is the default, and copyright is a limited exception justified only by a clear public benefit. The burden of proof thus rests squarely on those advocating for enclosure, a burden this paper argues cannot be met for raw GenAI outputs.

Current literature predominantly engages with this question through doc-trinal analyses of authorship, originality, and the idea-expression dichotomy, often taking existing legal frameworks for granted. This article makes a deliberate methodological choice: it prioritizes a foundational philosophical analysis over a purely doctrinal one. My central thesis is that GenAI's technical architecture (specifically its reliance on probabilistic inference) severs the causal chain between human intent and the work's final expressive form. The analysis therefore proceeds by testing whether copyright's traditional normative justifications can survive this fundamental break. To do so, I systematically evaluate the three pillars of copyright philosophy (utilitarian incentive, Lockean labor-desert, and Hegelian personality theory) against the reality of algorithmic expression.

This 'first principles' approach necessarily defines the scope and limitations of the inquiry. Consequently, this paper does not offer a detailed roadmap for resolving downstream doctrinal complexities, such as the mechanics of joint authorship between a human and a machine or the application of moral rights to a non-human entity's output. Instead, it argues that such questions are logically premature. Debating how to divide a right is fruitless if no right should exist in the first place. By establishing that raw GenAI outputs lack a coherent philosophical basis for protection, this article provides the necessary foundation upon which all subsequent doctrinal and policy discussions must be built.

The argument proceeds in four parts. First, I establish the technical and legal landscape by demonstrating how GenAI's probabilistic architecture severs the causal chain of human creative control (Section 2). This technical reality informs the subsequent mapping of current scholarly claims regarding GenAI authorship (Section 3) and diverse comparative legal responses (Section 4), highlighting the inadequacy of existing frameworks. Second, I systematically evaluate the dominant philosophical justifications for copyright law (utilitarianism, labor-desert theory, and personality theory) showing how they fail to provide a coherent rationale for protecting purely algorithmic outputs (Section 5). Third, I then refute the primary counter-argument that sophisticated, prompt-based originality constitutes a sufficient basis for authorship (Section 6). Fourth, I conclude that, given the comprehensive failure of these justifications, raw GenAI outputs must default to the public domain to prevent unprecedented enclosure and uphold copyright's instrumental purpose (Section 7).

## 2. Prompts, probability, and the vanishing grip of human creative control

GenAI has altered the landscape of creative production, presenting an unprecedented challenge to the foundational tenets of copyright law. It is fundamentally different to old tools that use to facilitate human expression such as a paintbrush, a word processor, or a camera. GenAI systems such as OpenAI's ChatGPT, image generators like Midjourney and Stable Diffu- sion, and Google's Gemini actively produce novel creative outputs. Large models such as GPT-4o can generate full essays or images from a single sen- tence, with user keystrokes representing ¡0.1 % of the tokens in the final work (OpenAI System Card, 2024), underscoring how little direct human control is required. This, compels us to confront a crucial philosophical question: should copyright law, a legal framework fundamentally rooted in the con- cept of human authorship, extend protection to works created autonomously or semi-autonomously by these sophisticated algorithms? In my view, the very mechanisms by which GenAI operates disrupt our understanding of creative control and expressive intent, thereby precluding their straightforward assimilation into existing copyright coverage.

The tension around mediated machine content generation is not entirely new. As early as 1965, the U.S. Register of Copyrights Abraham Kaminstein asked whether computer-generated works deserved protection, wondering if the machine "actually conceived and executed the traditional elements of authorship" (Miller, 1993, p. 1044) [1]. CONTU's 1979 report answered with a confident analogy: the computer was merely a "camera or typewriter," leaving authorship with the user (Samuelson, 1986, p. 1194) [2]. That analogy now rings hollow. Contemporary GenAI systems demand a more nuanced inquiry, as their core attributes of operational speed, functional autonomy, and technical opacity fundamentally distinguish them from mere instruments (Chesterman, 2021) [3]. This departure stems from their foundational architecture and training.

Through intensive pre-training on vast datasets, models built on architectures like the Transformer learn intricate patterns by using self-attention mechanisms to weigh the importance of different data points in relation to each other (Vaswani et al., 2017; Zhao et al., 2023) [4,5] (Figs. 1 and 2).

Crucially, GenAI models do not store copies of the training data in a sim- ple, retrievable sense. Instead, they develop an internal statistical representa- tion of the domain by learning probabilistic relationships between elements. When prompted by a user, the model leverages these learned relationships to generate new content. For text, this often involves sequentially predicting the most probable next word or 'token' (Zhao et al., 2023) [5]. For images and audio, a common technique is to start with random noise and progressively refine it into a coherent output that matches the prompt, a process known as diffusion (Ho, Jain, & Abbeel, 2020; Stability AI, 2024) [6,7]. This synthesis often occurs within a high-dimensional "latent space" before rendering the final output, a process distinct from a human artist drawing inspiration; the AI constructs new expressions based on statistical probabilities derived from its training.

User interaction with GenAI primarily occurs through prompts, which can range from simple textual descriptions to highly detailed instructions. The prompt serves as a catalyst and a guide, influencing the AI's probabilistic generation process. However, the relationship between the prompt and the resulting output is rarely deterministic. Many GenAI models exhibit what is often termed a "black box" nature, where the internal workings are not fully transparent, and identical prompts can produce materially different results due to inherent stochasticity. This unpredictability raises significant questions about the extent of human control over the final expressive elements, a control that copyright law traditionally exists to reward and protect.

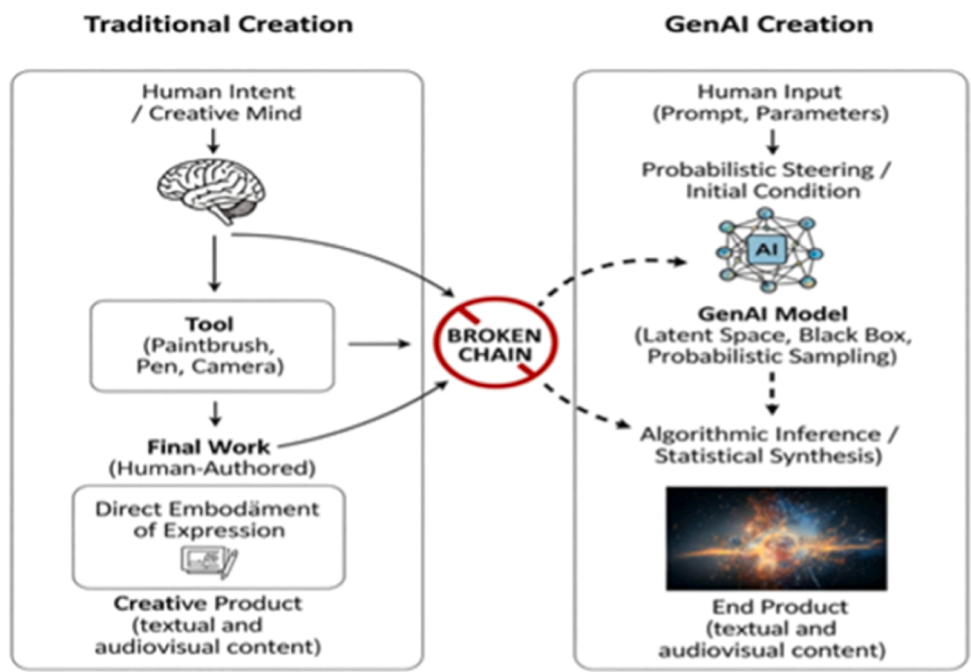

**Fig. 1.** The broken causal chain: human intent vs. algorithmic expression. This flowchart visually distinguishes traditional creative processes, where human intent directly shapes the final work, from GenAI creation, where probabilistic inference by the AI model severs the direct human-to-expression link, challenging traditional authorship.

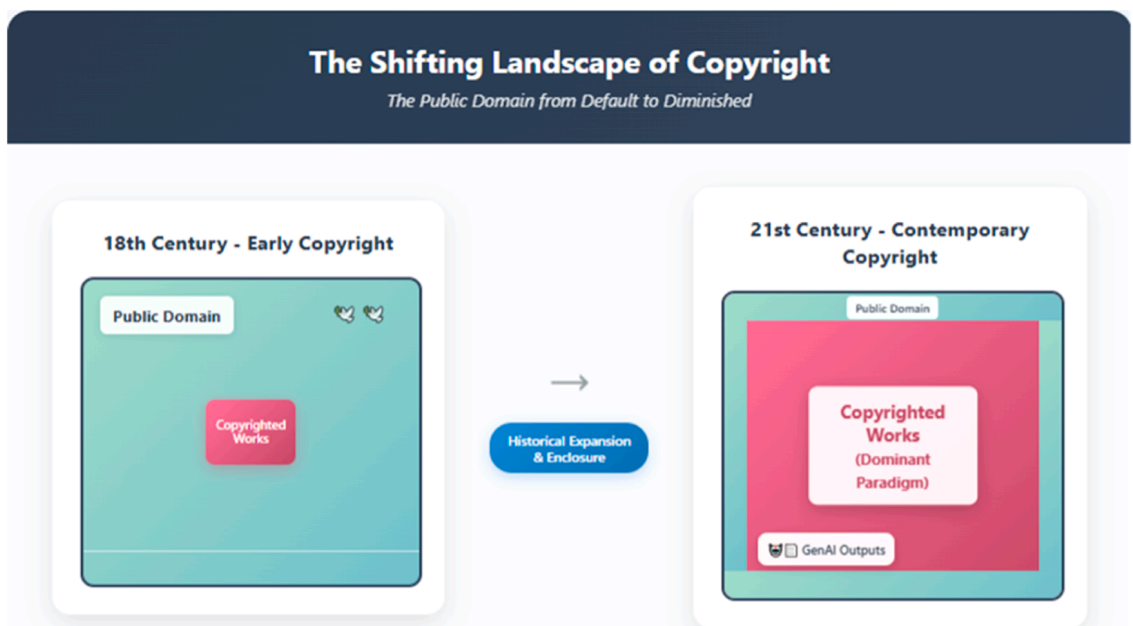

**Fig. 2.** The erosion of the public domain and the case for GenAI outputs as commons.
These visual traces the historical shift in copyright's scope, from a public domain default to a dominant paradigm of protected works, asserting that raw algorithmic outputs, lacking human authorship, inherently belong to the public commons.

Indeed, the prompt functions less like a direct instruction and more like an initial condition in a stochastic system. A single cue, such as "Seascape, Turner style," can yield an infinity of materially distinct canvases. Slight tweaks in seed values or temperature settings can dramatically alter the trajectory through the latent space, leading to divergent outputs. This inherent randomness (what computer scientists refer to as sampling noise) renders the process both opaque and difficult to replicate. (LeCun, Bengio, & Hinton, 2015) [8]. The probabilistic nature of these models, particularly those employing techniques like denoising diffusion, means that the user's input, while influential, does not equate to the direct, granular control over expression that underpins traditional notions of authorship. This is exemplified in leading models across different media: text-to-image systems like Stable Diffusion 3 use a Multimodal Diffusion Transformer to translate prompts into visual representations (Stability AI, 2024) [7], while audio generators like MusicGen use a similar Transformer-based process to convert text into discrete audio tokens that form a musical piece (Copet et al., 2023). This fundamental disconnect between human intent and algorithmic execution forms the crux of the philosophical challenge GenAI poses to copyright

## 3. Who owns the algorithmic brush? critiquing the claims for ownership

Scholarly debate has coalesced around four primary positions on who, if anyone, should own copyright in works produced with generative AI. Each camp offers its own premises about creativity, volition, and statutory design. However, the first three positions, which argue for some form of protection, ultimately fail because they cannot overcome the fundamental disconnect between human input and algorithmic expression—the broken causal chain at the heart of GenAI.

A first cohort treat GenAI as a sophisticated instrument rather than a creative agent. Burk (2020) likens it to a word processor and rejects the idea that we credit brushes or pencils with authorship (p. 266) [9]. Murray's (2024) "tools do not create" thesis reaches the same conclusion [10]: every protectable element must trace back to a person. Others lean on historical analogy: CONTU (1979) treated the mainframe as no more than an amped-up typewriter; Balganesh (2017) grounds authorship in human causation, rendering the machine legally invisible [11]. But these analogies break down precisely where they matter most: a word processor executes deterministic commands, while GenAI makes probabilistic leaps that sever the causal chain between human input and expressive output. The brush metaphor fails because no painter releases their brush to complete the painting based on its own statistical inferences.

Abbott and Rothman (2023) argue for "AI authorship" with legal surrogates holding the title, aiming to avoid what they call elaborate fictions (p. 1201) [12]. Salami (2020) supports full legal personhood for autonomous systems so that the entity performing the creative labor receives the right [13]. This position embraces the broken chain but proposes a remedy that multiplies rather than resolves the conceptual problems. The Ninth Circuit in Naruto v. Slater rejected non-human claimants, and the U.S. Copyright Office (2023) still requires a "human author [14,15]." Attaching authorship to a stochastic architecture that cannot bargain, license, or testify creates a legal fiction that serves no practical purpose. It would multiply orphan works, not foster progress, by vesting rights in a non-entity incapable of exercising them.

A pragmatic middle ground vests rights in the party that arranges for the AI's generation, typically the user. This approach finds a parallel in existing U.K. law, which deems the "person by whom the arrangements necessary for the creation of the work are undertaken" as the author of computer-generated works (Copyright, Designs and Patents Act 1988, 9(3)), and is supported by scholars who see prompt engineering as a sufficient exercise of creative control (Ginsburg & Budiardjo, 2019; Lee, 2025) [16]. Yet, this position falters on the requisite level of human control. Because the final output is the product of a probabilistic system, the user's prompt is merely an initial condition, not a direct act of creation A single seed adjustment can redirect a diffusion model toward an unforeseen aesthetic; predicting the final output often borders on the impossible (Ho, Jain, & Abbeel, 2020) [17]. If the human prompter cannot foresee or dictate the work's final expressive elements, the premise that they authored its expression grows tenuous, challenging the very notion of "sufficient control" that copyright traditionally rewards. This also highlights a critical gap in the literature: how granular must a user's intervention be to count as authorship, and how do we account for collective or layered authorship scenarios where trainers, model owners, and end-users each inject creative judgment at different stages? In other words, this approach mistakes influence for authorship and ignores the dispositive creative role played by the algorithm itself.

Finally, the only philosophically coherent position, and the one this article defends contends that raw GenAI outputs lie outside the statutory domain altogether. This conclusion follows directly from the failure of the preceding claims. This view is powerfully articulated by scholars like Carys Craig and Ian Kerr (2021), who argue that the very notion of an "AI author" is a "category mistake." They contend that authorship is a fundamentally relational and communicative human act, and therefore the "demise of the romantic author…should spell the death of the AI author" (p. 31) [18]., leaving the resulting works authorless and, by extension, in the public domain. Grimmelmann (2016) calls AI authorship "a good thing too" precisely because it channels creations into the commons [19]. Sun (2022) endorses a bifurcated regime: AI-assisted works may gain thin protection, but autonomously generated content should "redesign copyright" by defaulting to public domain status [20]. As this paper argues, without identifiable human volition and a direct causal link to the expression, protection lacks normative coherence [11, 21].

The failure of these doctrinal workarounds reveals a deeper problem: on what philosophical grounds, if any, should copyright protect works in which statistical inference supplies much of the expression and human input amounts to probabilistic steering. Most scholarship tinkers with statutory work-arounds or policy fixes without re-examining first principles. I therefore set out to probe the moral, utilitarian, and democratic justifications for copyright when creativity is co-produced by a model and its prompter. The aim is to test whether the classic link between authorship, incentive, and reward can survive the algorithmic muse.

## 4. Comparative legal responses to generative AI authorship

Before examining specific national approaches, it is essential to ground the discussion in the international copyright framework that shapes them. This global architecture, while not explicitly defining 'author,' is built upon a human-centric foundation that strongly supports this paper's thesis.

### 4.1. The international framework: Berne, TRIPS, and the human author

The cornerstone of international copyright law, the Berne Convention for the Protection of Literary and Artistic Works, establishes a baseline for protection among its signatories. Article 2(1) extends protection to "every production in the literary, scientific and artistic domain, whatever may be the mode or form of its expression." While the term "author" is not defined, and this has remained the case through all subsequent revisions including the Stockholm (1967) and Paris (1971) Acts, the entire structure of the Convention, particularly its provisions on moral rights (Article 6bis) and the term of protection based on the author's life (Article 7), presupposes a human creator. The concept of protecting an author's personality and granting rights for their lifetime plus a set number of years is incoherent when applied to a non-human, non-mortal algorithm. Furthermore, Article 2(8) explicitly excludes "news of the day or… miscellaneous facts" from protection, underscoring the Convention's focus on original expression over raw information, a distinction relevant to the statistical outputs of GenAI.

The WTO's Agreement on Trade-Related Aspects of Intellectual Property Rights (TRIPS) reinforces this human-centric norm. Article 9 (1) of TRIPS mandates that all member countries comply with the substantive provisions of the Berne Convention, thereby importing its implicit human authorship requirement into the global trade regime. While TRIPS introduced protection for computer programs as literary works, this was understood to protect the human-written code, not the outputs of the programs themselves. Thus, the foundational international treaties, while technologically neutral in their language, are philosophically human-centric. They establish a global default where copyright is a right for human creators, placing the burden of justification on any nation seeking to extend protection to non-human-authored works.

### 4.2. National approaches: a spectrum of human-centricity

Copyright authorities worldwide are clarifying their positions on the pro- tectability of AI-generated works, consistently reaffirming that human au- thorship is fundamental, albeit with notable jurisdictional variations. In the United States, copyright protection categorically requires human creativity. The constitutional basis for this principle stems from the Copyright Clause, limiting protection to works created by "Authors," understood as human beings (U.S. Constitution, art. I, § 8, cl. 8). The U.S. Copyright Office maintains an uncompromising stance, emphasizing that a copyrightable work must reflect human creative effort rather than mere operation of an automated process (U.S. Copyright Office, 2025, p. 10) [22].

This position is unequivocally crystallized by the landmark case of *Thaler v. Perlmutter*. The D.C. District Court, in its 2023 ruling, held that "United States copyright law protects only works of human creation" and that "human authorship is a bedrock requirement of copyright" [23]. Crucially, the court found that because Dr. Thaler "played no role in using the AI to generate the work," the AI output categorically failed the human-authorship test. On appeal, the D.C. Circuit unanimously affirmed this stance, flatly declaring that "the Copyright Act requires all eligible work to be authored… by a human being" and that "the Creativity Machine cannot be the recognized author of a copyrighted work" (Thaler v. Perlmutter, No 23–5148 (D.C. Cir. 2024)). This echoes earlier judicial pronouncements, such as the Ninth Circuit's holding in *Naruto v. Slater* (2018), which similarly rejected non-human authorship claims for photographs taken by a monkey [14].

This categorical approach extends beyond isolated AI-generated works. In the 2023 *Zarya of the Dawn* case, the Copyright Office partially cancelled a registration for a graphic novel created using Midjourney, explicitly distinguishing between the protectable human-authored text and arrangement versus the unprotectable AI-generated images. The Office's reasoning rigorously emphasized that the user's prompts constituted insufficient creative control over the visual outputs, as the specific visual features emerged from Midjourney's interpretation rather than human direction (U.S. Copyright Office, 2023, pp. 8–9). Similarly, the Office has rejected registration for Jason Allen's *Théâtre D'opéra Spatial*, despite over 600 prompt iterations, consistently maintaining that extensive prompting alone cannot transform an AI user into an author for copyright purposes (U.S. Copyright Office, 2024). Consequently, the Copyright Office maintains that mere prompting of generative models constitutes insufficient human control for authorship (U.S. Copyright Office, 2025, p. 26).

International jurisdictions broadly align with this human-centered stan-dard but differ in implementation and thresholds. For example, the United Kingdom takes a pragmatic legislative stance, attributing authorship to the "person by whom the arrangements necessary for the creation of the work are undertaken" (Copyright, Designs and Patents Act 1988, § 9 [3]) [24]. However, a recent UK Intellectual Property Office consultation expressed uncertainty over whether this statute adequately covers modern generative AI scenar- ios, leaving existing law unchanged pending further review (UK Intellectual Property Office, 2024, pp. 22–24) [25].

In the European Union, member states similarly prioritize human author- ship. The Court of Justice of the European Union (CJEU) has consistently held that originality requires the work to be the "author's own intellectual creation," reflecting "the personality of its author', and displaying his or her 'free and creative choices'. A standard that implies a human mind making free and creative choices (Cofemel, C-683/17) [26]. This human-centric foundation is now buttressed by a sophisticated and pragmatic regulatory framework detailed in a 2025 EUIPO report. The report clarifies that the EU's approach is a dual system, combining the Copyright in the Digital Single Market (CDSM) Directive with the new EU AI Act. The CDSM Directive establishes a legal basis for Text and Data Mining (TDM) but crucially provides rights holders with an opt-out mechanism under Article 4. The AI Act then makes this mechanism binding, legally obligating providers of general-purpose AI models to implement policies to respect these reservations of rights [EUIPO, 2025, p. 13] [27]. This legislative focus on regulating the input process, by giving rights holders control over ingestion, rather than attempting to define authorship of the output, implicitly reinforces this paper's core thesis: that the probabilistic nature of GenAI outputs makes them philosophically unsuited for traditional copyright, forcing lawmakers to seek solutions at the point of data training instead.

A 2024 policy questionnaire by the EU Council found general consensus that copyright protection attaches only when human involvement is "signifi- cant," explicitly excluding wholly autonomous AI-generated works (Council of the European Union, 2024, p. 4) [28]. Judicial decisions reflect this con- sensus: Czech courts recently refused authorship claims involving outputs from OpenAI's DALL-E system, reasoning that these lacked the requisite human creative input (Council of the European Union, 2024, p. 6).

Other countries, such as Australia, Canada, and India, also lean to- wards human authorship. Australia and Canada are engaged in ongoing consulta- tions to clarify their copyright laws concerning AI-generated works (Innova- tion, Science and Economic Development Canada, 2021, p. 12 [29]; Select Committee on Adopting Artificial Intelligence, 2024, 166) [30]. India's Copy- right Act of 1957 similarly defines the author of computer-generated works as "the person who causes the work to be created" (§ 2(d)(vi)) [31]. No- tably, an Indian copyright regis- tration that initially listed an AI tool as a co-author was later withdrawn, highlighting ongoing definitional challenges (Sarkar, 2021) [32]. Ja- pan's May 2024 guidelines determine copyrightability on a case-by-case basis, considering factors such as the amount and con- tent of user prompts, the number of generation attempts, user selection from mul- tiple outputs, and any subsequent human additions (Legal Subcommit- tee under the Copyright Subdivision of the Cultural Council, 2024, p. 17) [33].

In contrast, China presents a particularly nuanced and evolving doctrinal position, demonstrating a dynamic interplay between recog- nizing and rejecting copyright for AI-generated outputs based on the demonstrated human contribution. A notable Beijing Internet Court decision, *Li v. Liu* (2023), recognized copyright protection for an AI-generated image because the plaintiff's extensive selection from over 150 prompts and subsequent editing constituted sufficient intellectual achievements reflecting personalized human expression (Li v. Liu, 2023, Jing 0491 Min Chu No 11,279) [34]. The ruling was significant as it framed copyrightability not around the algorithm's autonomy but around demonstrable human curation. This pragmatic, process-oriented test established a pathway to protection by evaluating the entire creative workflow. The required human input was thus substantial and iterative, involving continuous refinement, selection, and adjustment, not merely a single command, creating a standard where copyrightability hinged on a demonstrable investment of personalized aesthetic judgment in guid- ing the AI's output.

However, the more recent ruling in *Feng Runjuan v. Kuashi Plastic* (Zhangjiagang Court, March 2025) [35], presents a crucial counter- point, demonstrating a higher bar for demonstrating human authorship and control. In this case, Ms. Feng, a designer, used Midjourney to generate images of a butterfly-shaped children's chair by entering a text prompt. After her design was commercialized without permission, she sued for copyright infringement. The court ultimately dismissed the case, explicitly ruling that the AI-generated images lacked sufficient "author-driven expressions" to qualify as original works. The court's reasoning offers critical insight into the substantive language of these rulings, The court stated that if content is primarily auto-generated by AI, it should not be copyrightable, as the AI, rather than the user, pri- marily determines the output. Ms. Feng failed to provide enough documentation to prove her sufficient human creative inputs in the creation process. The court emphasized the necessity of a verifiable creative process demonstrating the adjustment, selection, and embel- lishment of the original images by adding prompts and changing pa- rameters, and deliberate, individualized choices and substantial intellectual input over the visual expression elements, such as layout, proportion, perspective, arrangement, color, and lines." A critical factor was the "unpredictability," uncontrollability, and unrepeatability of the AI-generated content. Ms. Feng admitted she could not regenerate the same AI pictures as she posted earlier because of the randomness and uncertainty of the Midjourney AI service that she used. This inability strongly suggested to the court that it was the AI program making the design decisions, not the person providing the prompts, implying a lack of granular control and deliberate artistic choice by the human. The court held that the prompts used and shared by Plaintiff in this case, are merely ideas, and thus not copyrightable either, reinforcing the funda- mental idea-expression dichotomy.

The *Feng* ruling thus clarifies that while China's approach does tie outputs to human contribution, the threshold for demonstrating "suffi- cient" input is stringent. It necessitates a high degree of mastery over the AI tool, allowing for predictable and repeatable outcomes, or at least a clear, documented chain of creative decisions and refinements that led to the specific output. This judgment effectively recalibrates the pendulum in China, moving from a position of relative openness to a more demanding standard for proving human creative control, partic- ularly when dealing with AI tools characterized by significant randomness.

However, across most of these jurisdictions, courts and lawmakers implic- itly treat GenAI as an auxiliary instrument (analogous to a camera shutter at best) rather than as a co-author. This backdrop is crucial for the analysis that follows. If copyright's core philosophies already regard the public do- main as the default and exclusivity as an exception, then machine-generated works, standing even further removed from human personality or labor, face an even steeper justifi- catory hill. The lingering questions revolve around the precise threshold of human involvement, particularly in light of AI's probabilistic nature, which complicates the traditional direct link between human intent and expressive outcome.

## 5. Copyright as policy: the philosophical imperative for justification

Copyright, at its core, is not an inherent right but a legislative grant. A policy tool requiring robust justification. This fundamental premise, as emphasized by scholars like William Patry (2011), dictates that copy- right's scope and duration should be strictly limited to what is necessary to enrich the public domain [36]. This perspective is crucial when evaluating the copy- rightability of AI-generated content, as it places the burden of proof squarely on those who seek to extend protection to this new form of production.

Indeed, as Merges (2011) emphasizes, intellectual property rights require clear philosophical justification precisely because they are ex- ceptions to the default norm of open access and the public domain (Merges, 2011, pp. 3–5) [37]. This framing underscores the critical need to interrogate whether the traditional rationales (utilitarian incentive theory, Lockean labor-desert argu- ments, or personality-based justifi- cations) can coherently extend to genera- tive AI outputs.

*5.1. The utilitarian rationale*

Utilitarianism, long dominant in Anglo-American copyright philosophy, holds exclusivity justified only if it promotes the creation and dissemination of works for public benefit (Fisher, 2001, pp. 4–5) [38]. Without the prospect of exclusive rights and the economic rewards they enable, creators would supposedly lack motivation to invest time and effort in producing socially valuable works. Applying this to GenAI, one might argue that copyrighting AI outputs incentivizes the development and use of GenAI tools, leading to a greater volume of content. Yet, applying incentive logic to AI-generated outputs proves problematic. Whose incentives are we bolstering: the hu- man prompter whose creative input may be minimal and whose marginal contribution to expressive form is tenuous at best, or the developers of AI models, whose incentives may already be adequately met through other eco- nomic mechanisms? AI systems themselves, as algorithmic constructs devoid of subjective motivations, clearly require no incentivization to create. Thus, granting extensive rights for minimal human inputs risks a utilitarian mis- match, inflating protection beyond what is necessary or justified (Merges, 2011, pp. 139–145).

A more sophisticated utilitarian argument might focus not on the initial prompt but on the human labor involved in curating, refining, and post-editing AI-generated content. The prospect of copyright, it is argued, could motivate creators to transform raw outputs into polished works and, conversely, denying protection would create a chilling effect. This claim, however, misunderstands the nature of the incentive. A public default for raw outputs does not discourage human-AI collaboration; rather, it properly calibrates the reward to the human contribution alone. Furthermore, the incentive to refine an AI-generated work already exists through market demands and professional standards; an artist or writer seeks to produce a high-quality final product regardless. To grant full copyright for the combined output would therefore be a disproportionate reward, allowing a potentially minor human contribution to privatize the underlying, uncopyrightable machine-generated material. The more precise and utilitarian-sound solution is to recognize copyright only in the human author's original contributions (the specific edits, arrangements, or additions) treating the AI-generated content as pre-existing, public domain material. This approach, analogous to that for derivative works, properly rewards value-added human effort without creating a deadweight loss by enclosing the algorithmic portion.

A final utilitarian defense attempts to link the protection of outputs to the rights of upstream creators, arguing that output copyright is necessary to ensure those whose works were used for training are properly compensated. This argument, however, fundamentally misaligns the problem with the proposed solution. It attempts to solve the legitimate challenge of *input compensation* with the inappropriate and ineffective tool of *output rights*. Granting a downstream monopoly to the user of an AI system does nothing to guarantee that value flows back to the innumerable creators whose works were ingested by the model. Instead, it creates a perverse outcome: a windfall for the party who expended the least creative effort (the user), while leaving the original creators uncompensated. The challenge of remunerating upstream rights-holders is a critical policy question, but it demands direct, targeted solutions (such as statutory licensing schemes or industry-wide levies) that address the act of ingestion itself, not a misplaced property right in the resulting expression.

*5.2. The labor-desert rationale*

The labor-desert theory, drawing heavily on Lockean philosophy, posits that creators are entitled to the fruits of their labor by virtue of mixing their personal exertion with resources from the common, thereby removing them into private ownership (Hughes, 1988, pp. 296–300) [39]. While this principle resonates intuitively, its application to copyright has always been fraught, culminating in the widely famous U.S Supreme Court's rejection of the "sweat of the brow" doctrine in Feist Publications, Inc. v. Rural Telephone Service Co.(1991). Feist affirmed that mere effort insufficient; the labor must be creative to warrant protection.

The application of this theory to generative AI outputs, therefore, strains credulity on multiple fronts. If copyright rewards creative labor, whose labor precisely is embedded in a Midjourney-generated artwork? Is it that of the user, the model's engineers, or the innumerable creators whose works constitute the training data? Even if we focus solely on the user, the Lockean claim remains tenuous. While a user may invest significant effort in prompt engineering, parameter tuning, and iterative refinement, this labor is directed at steering a probabilistic system, not at directly fixing expression. The user's role is more analogous to that of a client commissioning an artist: they provide the concept and constraints, but the final expressive choices are made by another agent- in this case, the algorithm. The Lockean logic falters because the essential act of creation (the translation of an idea into a specific, expressive form) is not performed by the human. The chain of causation is broken by the model's latent space and its probabilistic sampling process. The pivotal act of expression occurs within a black-box statistical model, not within the deliberative agency of a human being.

Consequently, any property right derived from labor must attach only to what the human actually created: the prompts themselves, or the specific, separable modifications made in post-production. To allow this ancillary labor to justify ownership over the entire algorithmic output would be to grant a property right far exceeding the labor invested, violating the very proportionality inherent in Lockean theory and resurrecting the discredited "sweat of the brow" doctrine in a new, algorithmic guise.

*5.3. The personality rationale*

Personality theory, anchored in Hegelian and Kantian conceptions of au- thorship, protects works as outward expressions of the creator's inner self, identity, and will (Merges, 2011, pp. 68–101). From a Kantian perspective, this is not merely about rewarding effort; it is about recognizing that a work is an extension of the author's rational agency and autonomy. An author projects their will into the world, and copyright, particularly through the lens of moral rights (*droit moral*), protects this intimate, metaphysical bon between creator and creation. The theory presumes an unbroken nexus between human intent and the final expressive form, celebrating the work as a vessel for the author's unique persona.

Generative AI fundamentally severs this nexus. The output of a GenAI model emerges not from conscious reflection or lived experience but from algorithmically mediated statistical associations. Even a highly detailed prompt functions as an initial condition for a probabilistic process, not as a direct act of authorial will. The final expressive form, the specific sequence of words or arrangement of pixels, is the product of probabilistic sampling within a high-dimensional latent space, a process fundamentally alien to the direct embodiment of human personality.

A Kantian analysis would find this dispositive. The user causes a work to be created, but they do not author it in the sense of imbuing it with their rational will. The machine, as a non-rational entity, is a mere thing, a means to an end, incapable of the autonomous creative judgment that personality theory exists to protect. The final work, therefore, cannot be regarded as a true extension of the user's persona. It is a statistical artifact, not a vessel for human identity. The intimate bond between author and work, which is the very soul of personality theory, is broken.

This conclusion is not confined to the strong, personality-centric *droit d'auteur* tradition. The argument holds with equal, if not greater, force in jurisdictions that adopt a more pragmatic or economic approach to moral rights. In these systems, such as in the United Kingdom or under the limited Visual Artists Rights Act in the United States, moral rights function less as a metaphysical bond and more as a tool to protect the author's professional reputation and the integrity of the market for their work. This rationale also collapses when applied to a machine. An algorithm has no professional reputation to be harmed, no honor to be slighted, and no standing in an artistic community to protect. Therefore, whether viewed through a Hegelian lens of personal will or a pragmatic lens of economic reputation, the conclusion is the same: moral rights are fundamentally and inextricably linked to a human author, and the concept cannot be coherently stretched to cover the outputs of a non-human system.

Copyright, then, operates only where a compelling public rationale sur- vives rigorous scrutiny. Without a principled account of how utilitarian, labor-desert, or personality theories map onto probabilistic, machine-driven expression, extending statutory exclusivity to GenAI outputs risks overreach- ing the very justification for granting any monopoly at all. As Péter Mezei (2020) cautions, any attempt to fit these outputs into the existing framework requires "stretching" the fundamental concepts of copyright beyond their breaking point (p. 2) [40]. The safer orientation, one faithful to copyright's instrumental roots, is to leave algorithmic works in the public domain unless a demonstrable, meaningfully human creative contribution demands otherwise.

## 6. The limits of prompt-based originality in AI outputs

Copyright only rewards human-originated expression. The question is whether a prompt, not the model's latent computations, can supply that authorship. Sophisticated prompt engineering is the only serious candidate for grounding copyright in a generative-AI workflow, Proponents of copyright for AI outputs frequently argue that sophisticated prompt engineering consti- tutes sufficient human creativity to satisfy the authorship requirement. This perspective posits that the user's detailed guidance, iterative refinement, and specific creative choices, rather than the AI's autonomous generation, constitute the requisite human contribution.

This argument is not without merit. As Mark Lemley (2024) notes, "com- ing up with the right prompt to generate what you want will sometimes be an art form in itself," potentially satisfying the "modicum of creativity" standard [41]." The spectrum of prompts demonstrates this varying degree of human creative direction. Consider a minimal input like, "Create an image of a cat." This zero-shot artistic prompt yields a highly probabilistic output, relying almost entirely on the AI's default interpretations and training data, showcasing minimal human creative direction. In contrast, a complex Chain of Thought (CoT) prompt for a sophisticated art piece exemplifies signifi- cant human creative input: "I am a visionary artist developing a large-scale, multi--sensory installation.. My core concept explores the ephemeral nature of memory within digital landscapes, aiming to evoke a profound sense of nostalgia, loss, and fragmented beauty. Let's think step by step to construct this blueprint: First, analyze the core abstract concepts… Second, outline the emotional and thematic progression… Third, for Phase One specifically, detail the following elements, explaining the artistic rationale behind each choice… Finally, provide the complete conceptual blueprint…" (derived from Google, Prompting Guide 101, p. 7 principles) [42]. This layered instruction, specifying persona, thematic elements, emotional progression, and a structured, iterative design process, strongly implies a guiding human artistic vision that informs the AI's output.

Judicial support for prompt-based originality, albeit from specific jurisdic-tions, provides notable traction. Chinese courts, for instance, have recognized copyright in AI-generated works when substantial human creative input is present. In Lin Chen v. Hangzhou Gaosi Membrane Technology (2024), the Hangzhou Internet Court granted copyright for an AI-generated image, em- phasizing the plaintiff's extensive creative control through iterative prompt engineering and manual editing (e.g., over 30 prompts and subsequent Pho- toshop enhancements) [43]. Similarly, Li Yunkai v. Liu Yuanchun (2023) saw the Beijing Internet Court award copyright to a Stable Diffusion-generated image, specifically referring to the plaintiff's detailed prompts and iterative adjustments as reflecting aesthetic choices and personal judgment, thus meet- ing originality standards [44].

However, critical challenges persist for fully attributing authorship to the prompter under established copyright philosophy. As Lemley (2024) ar- gues, the AI's intrinsic "interpretation" and generation process mediate the prompter's intent, complicating the direct attribution of final expressive ele- ments solely to the human. The AI, operating through complex probabilistic algorithms and drawing upon vast training datasets, ultimately determines the specific arrangement of pixels, words, or sounds. This process differs fundamentally from an artist's direct control over paint or a writer's precise word selection. The inherent probabilistic nature often creates an "illusion of control" for the human user, challenging the traditional authorship nar- rative built on a direct, unmediated link between the author's creative mind and the resulting expression. Furthermore, even if a series of sophisticated prompts could be deemed a "work of authorship" on its own, any resulting copyright would likely be "thin." Such protection would cover only the spe- cific sequence and content of the prompts themselves, not the AI-generated output (without direct human editing, say through photoshop as highlighted in Chinese case law).

Consequently, copying the output, particularly if generated probabilisti-cally and not as a deterministic replication of the prompt sequence, might not infringe the copyright in the prompts. Finally, a crucial question remains whether the human's contribution to the output as a whole rises above the de minimis threshold. Under pre-vailing joint authorship principles, each con- tributor must provide independently copyrightable expression; mere ideas or general direction are insufficient (for instance U.S case S.O.S., Inc. v. Payday, Inc.) [45]. If the AI's expressive contribution is substantial, and the human's input is primarily limited to prompts or selecting from AI-generated options, the human's contribution to the final output may not meet the independent copyrightability standard required for authorship, even in a collaborative context (Ginsburg & Budiardjo) [46].

## 7. The public domain as constitutional default for GenAI

Copyright's starting point is the public domain. Statutory exclusivity is granted only when it can be shown, clearly and concretely, to advance public purposes (Fisher, 1988) [47]. That premise is rooted in the Statute of Anne and enshrined in the U.S. Constitution's mandate "To Promote the Progress of Science and useful Arts [48]". It frames copy-right not as a natural right of property but as an instrumental grant, justified only by its capacity to advance public welfare. When we ask whether GenAI outputs deserve protection, the inquiry therefore begins with a presumption of openness, not ownership. The onus rests on proponents of protection to rebut that presumption.

This binary framing of the issue (pitting full copyright against the public domain) is sometimes challenged by proposals for hybrid re-gimes, such as a neighboring right for the producer of the AI output or other forms of "thin" protection. While these alternative models are valid subjects for legislative debate, they are not interpretations of copyright law; they are proposals for new, *sui generis* rights. A neighboring right, for instance, traditionally protects the investment and organizational effort of an entrepreneur (like a music producer or broadcaster), not the original expression of an author. To propose such a right for GenAI outputs is to implicitly concede this paper's central claim: that these works lack the human authorship and originality required to qualify for copyright protection in the first place. For the sake of doctrinal integrity and principled clarity, these alternatives should be recognized as what they are: policy tools outside the philosophical boundaries of copyright.

Furthermore, these hybrid models are not a panacea for the problem of enclosure. While a "thin" right may seem less restrictive than a full copyright, it is still a new layer of exclusive control that encumbers the public domain without proper justification. It creates transaction costs, requires monitoring and enforcement, and can still lead to the creation of a "digital landfill" of legally radioactive works whose ownership status is unclear. The burden of proof must therefore remain on the proponents of any new right (thick or thin) to demonstrate a clear market failure that justifies this enclosure. Given that the incentives for developing and using GenAI are already robust, the case for even a limited property right is weak. The principled approach is to affirm the public domain as the default under copyright law, forcing any proposal for a new, alternative right to be rigorously justified on its own terms and weighed against the profound public benefit of a vibrant, open commons.

This starting point is not a mere academic preference; it is a corrective to a century of expansionist drift. Historically, copyright laws were far narrower than they are today. Early copyright in the United States, for in- stance, was primarily a regulatory tool focused on controlling the printing industry, not a broad property right over creative expression itself (Patry, 2011). The Copyright Act of 1790 initially protected only maps, charts, and books. Edward Samuels (1993, p. 163) meticulously documents the sub- sequent, significant expansion of copyrightable subject matter by the U.S. legislature throughout the 19th and 20th centuries, progressively enclosing historical prints (1802), musical compositions (1831), dramatic works (1856), photographs (1865), paintings, drawings, and sculptures (1870), lectures and motion pictures (1909), sound recordings (1971), pantomimes and choreographic works (1976), and computer programs (1980). This relentless expansion, Samuels (1993, p. 164) notes, has led to a "diminution in works that are treated as part of the public domain, to the point where there are few subject matter categories that are automatically considered as part of the public domain [49]."

James Boyle (2003, p. 140) famously characterizes this phenomenon as a "second enclosure movement," drawing a parallel to the historical enclosure of common lands in England. He argues that the open zones of knowledge and culture are shrinking as IP protection zones expand, questioning whether society is better off for it [50]. Boyle (2004, p. 1) views this expansion as hav- ing lost the "original principle of balance between knowledge, which should stay in the public domain free for all to use, and that which could be priva- tized [51]." Julie Cohen (2006, p. 121) similarly notes that IP is expanding "in length, breadth, depth, and strength," describing it as a "commodifica- tion" of culture and knowledge [52]. Lawrence Lessig (2006,) is skeptical of the need to privatize knowledge through copyright, suggesting the expan- sion is driven less by the "logic of incentives" and more by "the dynamics of political power [53]".

This historical perspective underscores that copyright, as it exists today, is the product of legislative choices and policy goals, not an inherent enti- tlement. It is an exception to the default state of free ac- cess. The trend, however, has reversed the initial assumption that in- tellectual creations were not protectable unless good cause was shown. Merges notes the contemporary mindset often defaults to asking, "why not protect a new form of intellectual creation?" This shift, often justi- fied by what Boyle (2004, p. 2) calls "over- stated incentive rhetoric" ("the more rights the better"), frequently lacks empirical evidence and constitutes "policy without balance" (Boyle, 2004, p. 2).

It is against this backdrop of doctrinal erosion that the challenge of gen- erative AI must be confronted. To grant copyright to the outputs of these systems would not be a minor extension of existing law but a radical break from its philosophical moorings. As established, the core justifications for copyright (utilitarianism, labor, and personality) disintegrate when applied to the probabilistic expression of an algo- rithm. The utilitarian incentive for the machine is zero, and the argu- ment for incentivizing the user is tenuous at best, risking a massive grant of rights for minimal creative input. The Lockean labor theory falters where the chain of causation between human effort and final expression is severed by a stochastic black box. And the Hegelian personality theory becomes incoherent when the work is not an extension of a human will, but the statistical artifact of a model's training data. If none of the classical rationales carries the necessary weight, the constitutional de- fault must prevail. The conclusion is therefore inescapable: raw gener- ative AI outputs belong in the public domain. To do otherwise would not only be philosophically unsound but would inflict at least two reasons:

## 8. Conclusion

This Article began by questioning the normative foundation for extending copyright to the outputs of generative AI. The analysis pre- sented leads to an inescapable conclusion: raw, machine-generated works lack the philosophical grounding necessary for protection and must reside in the public domain. To argue otherwise is to mistake the function of a sophisticated tool for the act of human creation and to ignore the fundamental principles upon which copyright law is built. The temptation to grant ownership over these novel creations is strong, but it is a temptation that sound legal policy must resist. The case against protection is rooted in the comprehensive failure of copyright's core justifications. The utilitarian rationale collapses under scrutiny: the machine requires no incentive, and the prompter's minimal effort is a thin reed upon which to hang a state-sanctioned monopoly. Lockean labor-desert theory falters where a probabilistic algorithm performs the dispositive expressive labor, severing the causal chain from human effort. Finally, personality theory, which views copyright as an exten- sion of the author's will, becomes meaningless when the 'creator' is a stochastic model incapable of possessing a will to extend. Without coherent justification from any of these foundational perspectives, the argument for protection dissolves. To ignore this philosophical vacuum and grant copyright to AI outputs would be an act of profound legal and social folly. It would trigger an enclosure of the digital commons on an unprecedented scale, creating a legal quag- mire of billions of orphan works whose ownership is ambiguous and whose use is chilled. This would not "Promote the Progress of Science and useful Arts," but actively stifle it, burying future creativity under a mountain of trans- action costs and legal uncertainty. It would represent a radical departure from copyright's instrumental purpose, transforming a limited, func- tional grant into a default property right for any and all machine- generated content.

This conclusion does not render human creativity obsolete, nor does it devalue works made with the assistance of AI. On the contrary, it clarifies the line. Where a human author engages in substantial, creative modification of an AI-generated output—through selection, arrange- ment, and alteration that meets the threshold of originality, copyright can and should protect that human contribution. The protection, how- ever, attaches to the human authorship, not the raw algorithmic output. The law must distinguish the artist who uses a digital tool from the operator who merely initiates a generative process.

Operationally, this distinction imposes a new evidentiary burden on those seeking protection for works incorporating GenAI. The dispositive question will no longer be *if* GenAI was used, but *how* it was used and to what extent the final work is a product of demonstrable human crea- tivity. Creators and their counsel must be prepared to document a sub- stantive transformative process: a record of iterative prompting that shows a clear creative trajectory, detailed logs of post-generation editing and arrangement, and a coherent articulation of the aesthetic judgments that shaped the final output from the raw algorithmic material. Courts, in turn, will be called upon to scrutinize these records, not to reward the 'sweat of the brow' in prompting, but to identify the specific, original contributions that are the proper subject of copyright. This approach transforms the legal inquiry into a forensic examination of the creative workflow, ensuring that protection attaches only to the human artist's discernible layer of expression, while leaving the underlying machine output in the public domain where it belongs

Ultimately, the challenge of generative AI forces copyright law to return to first principles. The public domain is not a void to be filled but the baseline of our intellectual ecosystem—the fertile ground from which all new creation springs. Affirming its dominion over the outputs of the algorithmic muse is not a failure of legal imagination but a reaf- firmation of copyright's core purpose. It ensures that the law remains a balanced instrument for promoting human creativity, not a mechanism for granting windfalls at the expense of the public good. \

**Funding**

Funded by Qatar University- High Impact Grant- QUHI-CLAW-25/26–775.

**Declaration of competing interest**

The author declares no known competing financial interests or personal relationships that could have appeared to influence the work reported in this paper.

**Data availability**

No data was used for the research described in the article.